\documentclass[aps,pra,twocolumn,showpacs,superscriptaddress,footinbib,preprintnumbers,longbibliography]{revtex4-2} 
\usepackage{graphicx}
\usepackage{dcolumn}
\usepackage{bm}
\usepackage{amssymb}
\usepackage{amsmath}
\usepackage{comment}
\usepackage{graphicx}
\usepackage{color}

\usepackage{MnSymbol}
\usepackage{slashed}
\usepackage[dvipsnames]{xcolor}
\usepackage{braket}
\usepackage{amssymb,amsmath,graphicx,color, quantikz}
\usepackage[export]{adjustbox}
\usepackage{braket}
\usepackage[colorlinks=true,citecolor=blue,linkcolor=red]{hyperref}
\usepackage{multirow}
\usepackage{rotating,booktabs}
\usepackage[verbose]{placeins}
\usepackage[caption=false]{subfig}
\usepackage{float}

\usepackage{soul}
\usepackage{umoline}

\newcommand{\Eq}[1]{Eq.~(\ref{#1})}

\newcommand{\Fig}[1]{Fig.~\ref{#1}}

\begin{document}

\preprint{UMD-PP-022-07}

\title{
Toward Quantum Computing Phase Diagrams of Gauge Theories
\\
with Thermal Pure Quantum States
}

\author{Zohreh Davoudi}
\email{davoudi@umd.edu}
\affiliation{Maryland Center for Fundamental Physics and Department of Physics, University of Maryland, College Park, MD 20742, USA}
\affiliation{Institute for Robust Quantum Simulation, University of Maryland, College Park, Maryland 20742, USA}

\author{Niklas Mueller}
\email{niklasmu@umd.edu}
\affiliation{Maryland Center for Fundamental Physics and Department of Physics, University of Maryland, College Park, MD 20742, USA}
\affiliation{Joint Quantum Institute, NIST/University of Maryland, College Park, MD 20742, USA}

\author{Connor Powers}
\email{cdpowers@umd.edu} \thanks{(corresponding author)}
\affiliation{Maryland Center for Fundamental Physics and Department of Physics, University of Maryland, College Park, MD 20742, USA}
\affiliation{Institute for Robust Quantum Simulation, University of Maryland, College Park, Maryland 20742, USA}

\begin{abstract}
The phase diagram of strong interactions in nature at finite temperature and chemical potential remains largely unexplored theoretically due to inadequacy of Monte-Carlo-based computational techniques in overcoming a sign problem. Quantum computing offers a sign-problem-free approach but evaluating thermal expectation values is generally resource intensive on quantum computers. To facilitate thermodynamic studies of gauge theories, we propose a generalization of thermal-pure-quantum-state formulation of statistical mechanics applied to constrained gauge-theory dynamics, and numerically demonstrate that the phase diagram of a simple low-dimensional gauge theory is robustly determined using this approach, including mapping a chiral phase transition in the model at finite temperature and chemical potential. Quantum algorithms, resource requirements, and algorithmic and hardware error analysis are further discussed to motivate future implementations. Thermal pure quantum states, therefore, may present a suitable candidate for efficient thermal-state preparation in gauge theories in the era of quantum computing.
\end{abstract}
\maketitle

\noindent
\textit{Introduction.}
The notorious sign problem in classical Monte-Carlo-sampling techniques has slowed down progress in addressing a range of problems in the domain of nuclear and high-energy physics. These include \emph{ab initio} nuclear many-body calculations of nucleonic matter including large-mass atomic nuclei~\cite{carlson2015quantum,lahde2015nuclear,roscher2014fermi}, and lattice Quantum Chromodynamics (QCD) calculation at finite baryonic density and chemical potential~\cite{de2010simulating,ding2015thermodynamics, soltz2015lattice, aarts2016introductory, ratti2018lattice, bazavov2019hot}, pertinent to `critical endpoint' searches in the Beam Energy Scan at the Relativistic Heavy Ion Collider~\cite{cebra2014studying,petersen2017beam}, and essential in illuminating exotic QCD phases potentially present in the interior of neutron and quark stars~\cite{Kogut:2004su,stephanov2006qcd, fukushima2010phase, guenther2021overview}. Attempts to alleviate sign problems are numerous and include reweighting~\cite{kieu1994monte,fodor2002new, fodor2002lattice}, Majorana~\cite{li2015solving,li2016majorana} and Meron Cluster algorithms~\cite{chandrasekharan1999meron}, stochastic quantization and complex Langevin dynamics~\cite{aarts2008stochastic,aarts2009can,aarts2013complex,aarts2010complex, kogut2019applying,berger2021complex}, Taylor expansion around vanishing chemical potential~\cite{de2002qcd,ejiri2004study,karsch2011towards,bonati2018curvature,bazavov2019chiral}, analytic continuation from imaginary chemical potential~\cite{bellwied2015qcd,bonati2015curvature,borsanyi2020qcd}, and path deformation and complexification~\cite{cristoforetti2012new,cristoforetti2013monte, bacsar2017going,alexandru2017tempered,alexandru2020complex}. Monte-Carlo technique can be avoided altogether using Hamiltonian-simulation methods such as Tensor Networks~\cite{banuls2015thermal,banuls2016chiral,banuls2017density,buyens2016hamiltonian} and quantum-simulation-inspired algorithms~\cite{czajka2021quantum}. Nonetheless, despite this progress, a universal approach to overcome the sign problem in lattice-QCD calculations is not known.

Digital quantum computers, as well as analog quantum simulators, offer a sign-problem-free approach to computing ground, excited, thermal, and non-equilibrium states of quantum many-body systems including lattice gauge theories (LGTs)~\cite{banerjee2012atomic,zohar2013simulating, zohar2015quantum,mildenberger2022probing, yang2016analog,zache2018quantum, davoudi2020towards,surace2020lattice, luo2020framework, andrade2022engineering, klco2018quantum, lu2019simulations, barbiero2019coupling, Chakraborty:2020uhf, shaw2020quantum, stryker2021shearing, davoudi2021toward, homeier2021z, Pederiva:2021tcd, Rajput:2021khs,martinez2016real,nguyen2022digital, mil2020scalable,zhou2021thermalization,byrnes2006simulating, zohar2013quantum,zohar2013cold,tagliacozzo2013simulation,de2021quantum,klco20202,atas20212,rahman20212,haase2021resource,kan2021lattice,davoudi2021search,Ciavarella:2021nmj,Alam:2021uuq,Paulson:2020zjd,halimeh2022gauge,Ciavarella:2021lel,lamm2019general,cohen2021quantum, gonzalez2022hardware, atas2022real,farrell2022preparations,Murairi:2022zdg,banuls2020simulating,Klco:2021lap,davoudi2022quantum}. However, because quantum devices naturally implement pure states, quantum computing thermal, i.e.,~mixed, states is a challenging task. An obvious possibility, thermalizing a non-equilibrium state through unitary time evolution~\cite{zhou2021thermalization,mueller2022thermalization}, may be costly on near-term quantum devices due to long equilibration times involved. Other methods for quantum computing thermal expectation values exist, which generally fall into two categories.  One is sampling easily-preparable pure states according to thermal statistics~\cite{Brandao:2016mfe,lu2021algorithms,motta2020determining,sun2021quantum,Temme:2009wa,kastoryano2016quantum,yung2012quantum}, but poor scaling of required samples with system size can lead to prohibitive resource requirements. Another is preparing  full (mixed) thermal states on the quantum computer, often relying on  resource-intensive algorithms such as quantum phase estimation and incurring significant ancillary-qubit requirements~\cite{bilgin2010preparing,poulin2009sampling,riera2012thermalization,zhu2020generation,chowdhury2016quantum,terhal2000problem}. Hybrid classical-quantum algorithms for near-term applications~\cite{lu2021algorithms}, such as those relying on variational methods~\cite{verdon2019quantum,chowdhury2020variational,wu2019variational,zhu2020generation,sagastizabal2021variational,wang2021variational,gomes2021adaptive,warren2022adaptive} or utilizing Jarzynski's equality~\cite{jarzynski1997nonequilibrium,schuckert2022probing,bassman2021computing} are also explored in recent years.
\begin{figure}[t!]
    \centering
    \includegraphics[scale=0.645]{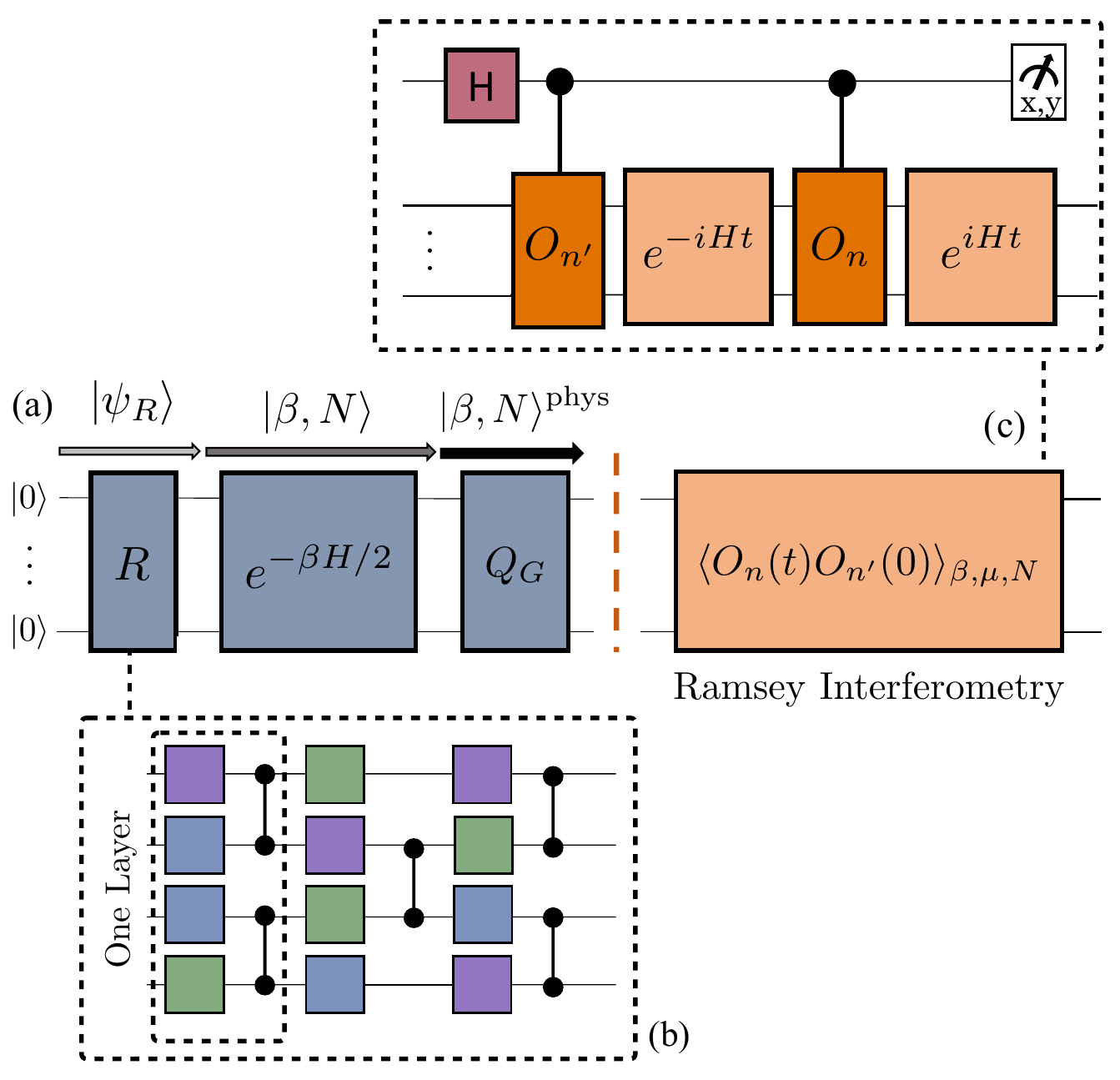}
    \caption{Schematic overview of the PTPQ-state preparation. The colored squares in the $R$ circuit denote randomly-chosen single-qubit rotations around $\mathsf{X}$ and $\mathsf{Y}$ axes of the Bloch sphere by angle $\frac{\pi}{2}$ and the $\mathsf{T}$ gates. The entangling gates are controlled-$\mathsf{Z}$ rotations. The two consecutive single-qubit gates on each qubit are constrained to be different in this construction. Entangling gates in each layer only act on adjacent qubits and this pattern continues, see Ref.~\cite{richter2021simulating} for details. $\mathsf{H}$ is a Hadamard gate. 
    }
    \label{fig:circuitoverview}
\end{figure}

A promising near-term perspective is to resort to the thermal-pure-quantum-(TPQ-) state formulation of statistical mechanics~\cite{sugiura2013canonical, sugiura2012thermal,hyuga2014thermal} to obtain thermal expectation values using only a single properly-prepared pure state in the thermodynamic limit. This approach offers a promising path to simulating quantum systems at finite temperature and chemical potential on quantum computers~\cite{powers2021exploring,coopmans2022predicting}. Canonical TPQ states, in particular, are formed by imaginary-time evolving pure states drawn from an ensemble of Haar-random states. Thermal expectation values of mechanical observables with TPQ states converge rapidly to the values within the standard (ensemble) formulation of statistical mechanics, improving exponentially with system size. In fact, as is shown in Ref.~\cite{sugiura2013canonical}, for sufficiently large systems often a single TPQ state suffices. However, in gauge theories, local symmetries put constraints on the allowed Hilbert space so that the na\"{i}ve TPQ-state approach cannot be applied. In this manuscript, a protocol for explicitly constructing TPQ states in the physical subspace of gauge theories will be introduced by penalizing non-physical components of the random pure state as they evolve in imaginary time.
 
By accurately obtaining, via a numerical simulation, the phase diagram of a simple gauge theory ($Z_2$ LGT in $1+1$ D with matter) at finite temperature and chemical potential, we demonstrate the utility of TPQ-state approach in studying thermodynamics of gauge theories for the first time. Aiming at quantum-computing applications, associated quantum algorithms, quantum-resource requirements, and robustness to algorithmic and hardware errors are further studied. The results indicate that the TPQ-state approach may be a suitable candidate for efficient phase-diagram studies of QCD in the future.

\vspace{0.1 cm}

\noindent
\textit{Thermal Pure Quantum States for Gauge Theories.} Canonical TPQ states are defined as~\cite{sugiura2013canonical}
\begin{equation}
\label{eq:defTPQ}
    |\beta,N \rangle \equiv {e^{-\frac{\beta }{2}H}}{}|\psi_R\rangle\,,
\end{equation}
with $\beta$ being the inverse temperature, $N$ the number of degrees of freedom of the (discrete) system, and $| \psi_R \rangle $ a Haar-random state. TPQ states approximate thermal
expectation values of `mechanical' operators, i.e.,~those that are low-degree polynomials of local operators, via
\begin{align}
    \langle O \rangle_\beta \approx \frac{\llangle  \, \langle \beta,N | O | \beta,N \rangle \, \rrangle_r}{\llangle  \, \langle \beta,N | \beta,N \rangle \, \rrangle_r}\,,
\end{align}
with exponential convergence in the system size (as well as in inverse temperature), see Supplemental Material~\cite{SM} for details.
While a single TPQ state suffices as $N \to \infty$, for faster convergence at finite $N$, a stochastic average over $r$ TPQ states can be preformed,  denoted by $\llangle \cdot \rrangle_r$ in the formula.

In gauge theories, $|\Phi_R\rangle$ may be unphysical, in which case \Eq{eq:defTPQ} will not reproduce physical thermal observables. While this issue can be avoided by eliminating the gauge-field degrees of freedom with certain boundary conditions in 1+1 D, such a strategy is not generally applicable. Therefore, we propose `physical' thermal pure quantum (PTPQ) states 
\begin{align}
\label{eq:physdefTPQ}
|\beta,N \rangle^{\rm phys} \equiv {e^{-\frac{\beta }{2}\tilde{H}}}|\Psi_R\rangle\,,
\end{align}
by adding a term to the Hamiltonian, $\Tilde{H}\equiv H+ \sum_n f(G_n)$ where $G_n$ are Gauss's law operators at site $n$, that is $[H,G_n]=0$. The function $f$ is chosen such that unphysical components of the state as it evolves in imaginary time are penalized in energy. Such an approach is customary in the context of enforcing Gauss's law in analog and digital quantum simulation of gauge theories, and can be applied to both Abelian and non-Abelian cases~\cite{banerjee2012atomic,hauke2013quantum,banerjee2013atomic,marcos2013superconducting,halimeh2021gauge,halimeh2022gauge,van2021suppressing,mathew2022protecting}.

A circuit to prepare PTPQ states on quantum computers is illustrated in \Fig{fig:circuitoverview}. First, a random circuit $R$, consisting of layers of single-qubit gates and entangling two-qubit gates, is used to prepare an approximate Haar-random state. Various designs are suggested for such task with studied performance~\cite{weinstein2008parameters}, and we adopt the efficient implementation of Ref.~\cite{richter2021simulating}. This random circuit is followed by a non-unitary operator $e^{-\beta H/2}$ acting upon the resulting random state to produce a standard canonical TPQ state. Gauss's law is enforced through action with $Q_G\equiv e^{-\frac{\beta}{2} \sum_n f(G_n)}$, the circuit implementation of which depends on the $f$ chosen, see below for the example of $Z_2$ LGT in 1+1 D. These elements will be further studied in the following.

\vspace{0.1 cm}

\noindent
\textit{Thermal chiral phase diagram of $Z_2^{1+1}$.} 
The model that will be studied in the following to demonstrate the value of the TPQ-state approach in gauge theories is $Z_2$ LGT in 1+1 D coupled to staggered fermions ($Z_2^{1+1}$). This model is sufficiently simple to allow numerical verifications on classical computers, while it still exhibits a non-trivial phase diagram which is aimed to be reproduced by quantum simulation. The Hamiltonian of the model is
\begin{equation}
\begin{aligned}\label{eq:full_Z2}
    H=  \frac{1}{2a}\sum_{n=0}^{
    N-2}(c^\dagger_{\scriptsize n} \Tilde{\sigma}^z_{\scriptsize n} c_{\scriptsize n+1} +{\rm h.c.})+m\sum_{n=0}^{N-1} (-1)^n c^\dagger_n c_n - \epsilon \sum_{n=0}^{N-2} \Tilde{\sigma}^x_n\,,
\end{aligned}
\end{equation}
where $c^\dagger_n$ ($c_n$) is fermionic creation (annihilation) operator, and $\bar{\sigma}^z_n$ and $\bar{\sigma}^x_n$ are Pauli spin operators realizing the $Z_2$ link and electric field operators, respectively. Open boundary conditions are considered throughout, and generalization to other boundary conditions is straightforward. $N$, $a$, $m$, and $\epsilon$ are fermionic lattice size, lattice spacing, fermion mass, and electric-field strength, respectively. Gauss's law operator $G_n\equiv \Tilde{\sigma}^x_n \Tilde{\sigma}^x_{n-1}e^{i \pi \left[ c^\dagger_n c_n + ((-1)^n -1)/{2} \right]}$ defines the physical subspace of the theory via the relation $G_n |\Psi\rangle^{\rm phys} = |\Psi\rangle^{\rm phys}$. For all the results shown, we set $a=1$, $m=1/2$, and $\epsilon/m=1$.
\begin{figure}[t]
\begin{centering}
\includegraphics[scale=0.535]{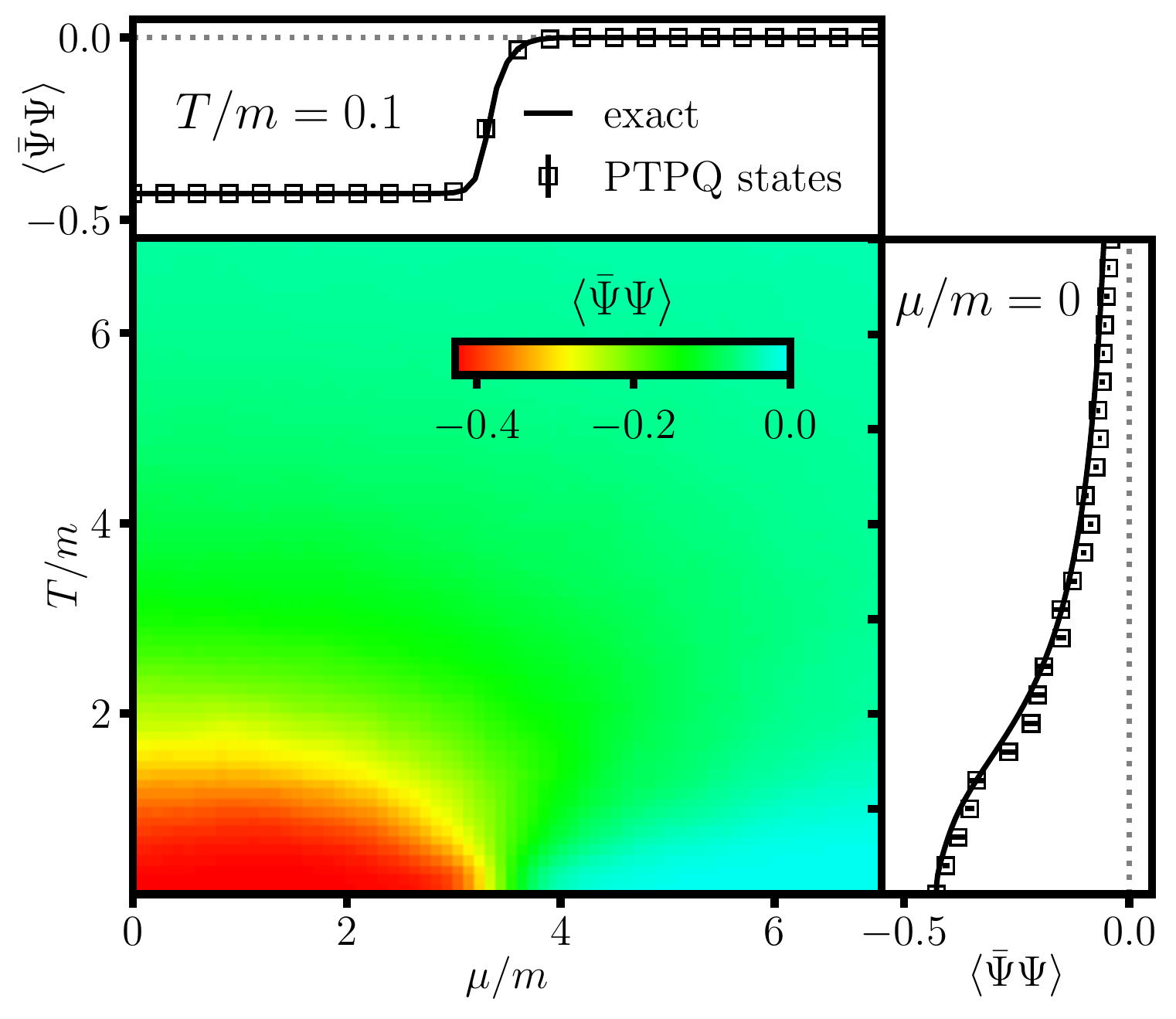}
\end{centering}
\caption{Chiral phase diagram for the $Z_2$ LGT with fermions in $1+1$ D with $N=6$ and $\lambda/m=4$, calculated with PTPQ states with random-circuit depth $d=20$ and averaged over $r=10$ PTPQ-state realizations. Error bars denote statistical uncertainty from a finite PTPQ sample.}
\label{fig:chiral_plots}
\end{figure}

Using PTQP states with $f(G_n)=\lambda (1-G_n)$ for large $\lambda$, the chiral phase diagram of $Z_2^{1+1}$ at finite temperature $T$ and chemical potential $\mu$ can be computed. A non-zero chemical potential can be accounted for by modifying the Hamiltonian, $H\rightarrow H - \mu\sum_{n=0}^{N-1} c^\dagger_n c_n$. The order parameter to be evaluated is the chiral condensate,
\begin{equation}
    \langle \Bar{\Psi}\Psi\rangle \equiv \frac{1}{N}\langle\sum_{n=0}^{N-1}(-1)^n c^\dagger_n c_n \rangle\,,
    \label{eq:chiral_condensate}
\end{equation}
with $\langle \Bar{\Psi}\Psi\rangle \neq 0$ ($=0$) in the chiral-symmetry broken (symmetric) phase. In QCD, chiral symmetry is spontaneously broken at small T~\footnote{In addition to a small explicit breaking due to finite bare quark masses.} but restored at large $T$~\cite{kogut1982scales,shuryak1993chiral,gottlieb1987estimating,morones2017chiral,karsch2002lattice}. Chiral symmetry breaking is also observed in 1+1 D Abelian gauge theories~\cite{hamer1982massive,melnikov2000lattice,frank2020emergence}. For $Z_2^{1+1}$ with staggered fermions, chiral symmetry restoration can be understood intuitively. At asymptotic $\mu \gg m,\epsilon$, the state with all fermion sites occupied dominates the thermal Gibbs ensemble and has $ \langle \Bar{\Psi}\Psi\rangle=0$ according to \Eq{eq:chiral_condensate}, while at $T \to \infty$, the state is an equal admixture of all basis states, giving rise to a net vanishing condensate.

Figure~\ref{fig:chiral_plots} depicts the phase diagram using PTPQ states with random-circuit depth $d=20$ and averaged over $r=10$ PTPQ-state realizations. All results are classically computed using exact diagonalization on a system of $N=6$ sites corresponding to 11 qubits~\cite{SM}. Chiral symmetry is observed to be restored at high $T$ and $\mu$ and is broken otherwise. Despite the small system size, the transition at $T\rightarrow0$ is visibly discontinuous, indicating that a `true' phase transition is expected in the infinite-volume limit, while only a smooth crossover is observed at $\mu\rightarrow 0$.
Top and right panels of \Fig{fig:chiral_plots} show a comparison between exact results and PTPQ states. Reasonable agreement is observed, and the accuracy is expected to improve with increasing system size, see Supplemental Material for details~\cite{SM}. Furthermore, as demonstrated in Supplemental Material~\cite{SM}, the standard TPQ states do no reproduce the correct phase diagram unless promoted to PTPQ states.

In contrast, regardless of the use of PTPQ states, detecting the confinement-deconfinement
transition, where one computes non-local observables, e.g., string tension, is more challenging on small systems due to finite-size effects, as discussed in Supplemental Material~\cite{SM}.

\begin{figure}[t]
\begin{centering}
\includegraphics[scale=0.465]{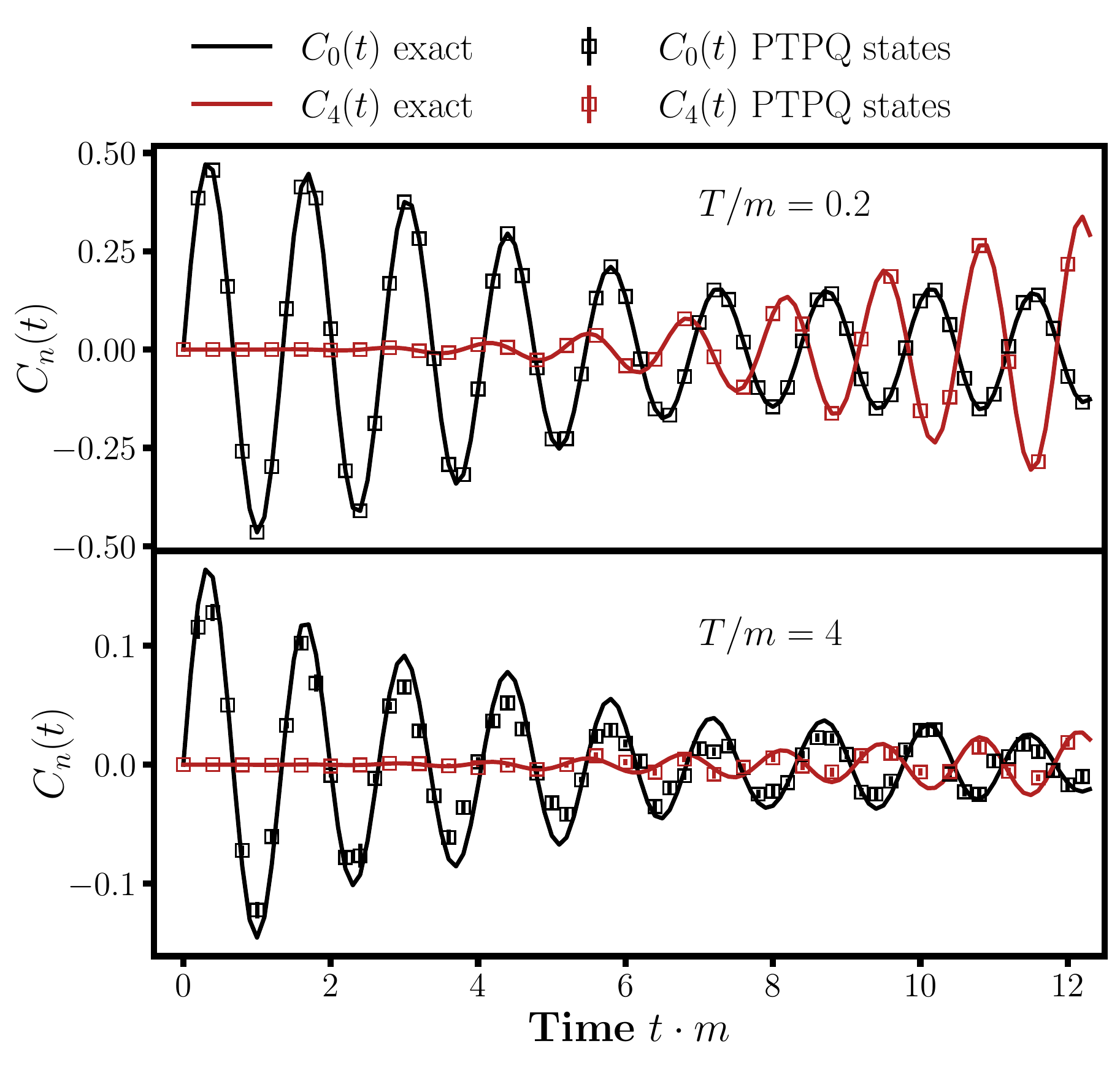}
\end{centering}
\caption{Non-equal-time correlation functions $C_0(t)$ and $C_4(t)$ with $N=6$, $\mu/m=0.5$, and $\lambda/m=4$. Data are shown at $T/m=0.2$ with $r=1$ and at $T/m=4$ with $r=20$.}
\label{fig:nonequal_time_correlators}
\end{figure}
\begin{figure*}[t]
\begin{centering}
\includegraphics[scale=0.655]{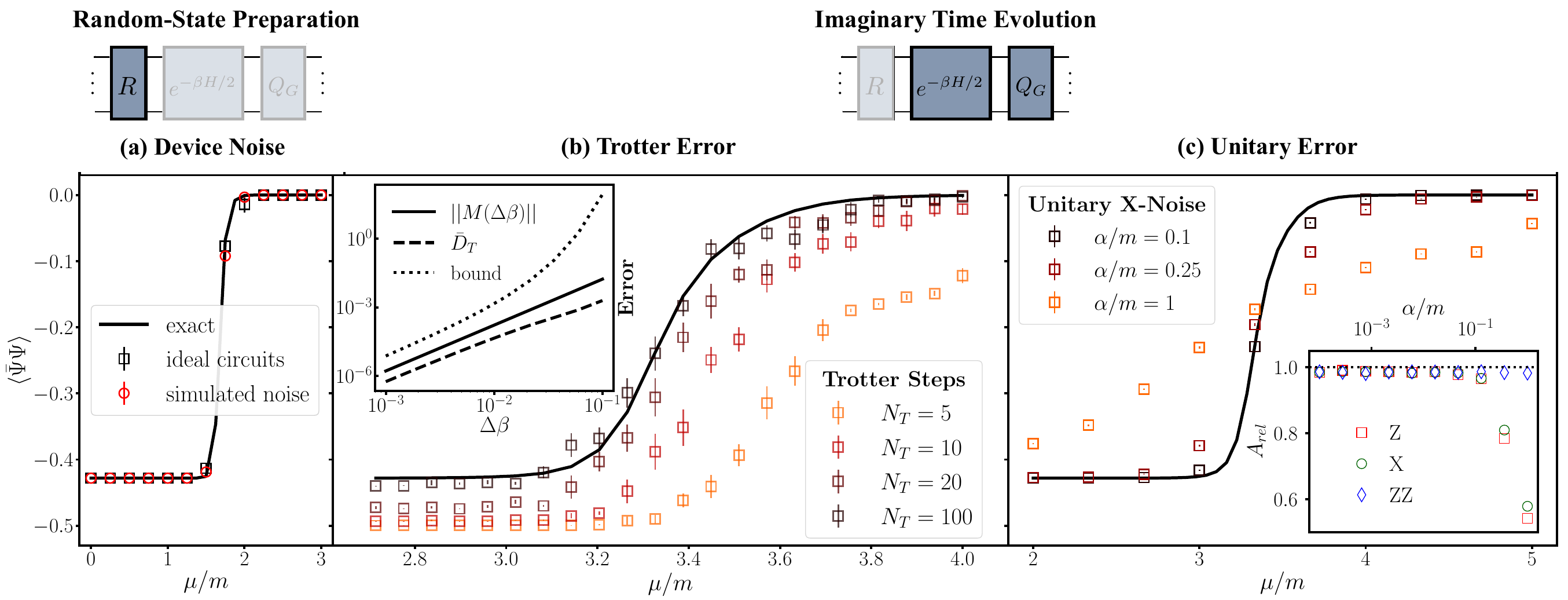} 
\end{centering}
\caption{
(a) Chiral condensate $\langle \bar{\Psi} \Psi \rangle$ as a function of $\mu/m$, including simulated device noise~\cite{Qiskit} on the Haar-random state-preparation circuit shown in \Fig{fig:circuitoverview}(b). (b) Trotterization error in evaluating $\langle \bar{\Psi}\Psi\rangle$ as a function of $\mu/m$ for various numbers of Trotter steps $N_T$. Inset: First-order product-formula Trotter error bound from Ref.~\cite{childs2021theory} as a function of $\Delta \beta$ for $N=6$ at $\mu/m=4$. Explicitly calculated multiplicative Trotter error $||M(\Delta \beta)||$ is plotted for comparison, as well as the mean trace distance $\bar{D}_{Tr}$ between $e^{-\Delta \beta H}|\Psi_R\rangle$ and $S(\Delta \beta)|\Psi_R\rangle$.
(c) Unitary errors in $\langle \bar{\Psi}\Psi\rangle$ as a function of $\mu/ m$ calculated with PTPQ states with introduced unitary Z errors at various relative Hamiltonian error strengths $\alpha/m$. Inset: Relative accuracy $A_{\rm rel}$ as a function of relative Hamiltonian error-term strength $\alpha/m$ at $\mu/m=3$ when various types of unitary error are added to the imaginary-time evolution. All results are obtained for $d=20$, $r=10$, $\lambda/m=8$, $T/m=0.1$, and averaged over results from 50 randomly generated error terms at each value of $\alpha/m$.}
\label{fig:error_plots}
\end{figure*}
Thermal non-equal-time correlation functions can also be evaluated using the PTPQ-state protocol. These quantities are challenging to access with classical Monte-Carlo techniques even at vanishing chemical potential. As an example, we focus on the thermal current-current correlator,
\begin{align}\label{eq:currcurr}
C_n(t) \equiv \langle [j_n(t), j_0(0) ]\rangle_{\beta}\,,
\end{align}
where $j_n(t) \equiv e^{iHt} j_ne^{-iHt}$, $j_n \equiv \frac{i}{2a}[\sigma^+_n \tilde{\sigma}^z_n \sigma^-_{n+1} - \rm{h.c.}]$ is the fermionic current operator, and $\langle \cdot \rangle_{\beta}$ denotes the thermal expectation value at finite $\beta$. 
A quantum circuit for evaluating $C_n(t)$ is presented in \Fig{fig:circuitoverview}(c) and involves a standard Ramsey interferometry scheme~\cite{somma2002simulating,knap2013probing,yao2016interferometric,tonielli2020ramsey}, requiring an ancilla in a Hadamard superposition controlling $j_n(t)$ and $j_0(0)$, with the PTPQ-state input. The results of this procedure (classically computed) are shown in \Fig{fig:nonequal_time_correlators} for $N=6$, $T/m=0.2$, and $\mu/m=0.5$. The exact results, obtained with exact diagonalization, are compared with a PTPQ computation using $r=1$ and $r=20$ samples for lower and higher temperatures, respectively, and good agreement is observed. Since real-time evolution is natural for a quantum computer, PTPQ states are expected to be an important tool to compute e.g., transport quantities like conductivity or viscosity. For example, one can use \Eq{eq:currcurr} to compute thermal conductivity, $\sigma_{\beta,\mu} \sim \lim\limits_{\omega\rightarrow 0 } \frac{a}{\omega} \int dt \, e^{i\omega t} \sum_n C_n(t)$, which nonetheless requires long evolution times and large system sizes to be obtained accurately.

\vspace{0.1 cm}

\noindent
\textit{Robustness to Quantum-Circuit Implementation Errors.}
To investigate inaccuracies occurring in the quantum-circuit implementation of the PTPQ-state preparation in $Z_2^{1+1}$, we recall that this preparation consists of two parts, generation of a Haar-random state, followed by imaginary-time evolution with $H + \lambda \sum_n(1-G_n)$. 

First, by employing an exemplary current noise model, the effect of device noise in generating a Haar-random state can be investigated using the circuit described in Fig.~\ref{fig:circuitoverview}(b). Shown in \Fig{fig:error_plots}(a) is the chiral transition of an ideal device versus IBM's `Sydney' device~\cite{Qiskit} (simulator) when noise is present in the Haar-random-state preparation, while the imaginary-time evolution is performed exactly (shot noise is not considered). Data corresponds to a circuit depth of $d=20$ and $r=10$ PTPQ realizations for a system of $N=6$ fermionic sites. As is seen, device errors included in current simulators have a negligible effect during this part of the algorithm. Whether features like phase transitions remain robust with respect to the actual device error will need to be tested on the available quantum hardware in future studies. 

The second part of the algorithm is imaginary-time evolution. Here, an error analysis will specifically depend on the chosen implementation strategy and device. Among the common near-term approaches include the Quantum Imaginary Time Evolution (QITE) algorithm~\cite{motta2020determining} and dilated operator approaches~\cite{gingrich2004non,turro2022imaginary}. Both approaches employ Trotterization for imaginary-time evolution, hence it is beneficial to investigate the Trotterization error in the PTPQ-state preparation. 

For simplicity, we consider Trotterization via a first-order product formula of the form
\begin{align}
\label{eq:Trotterform}
    S(\beta) \equiv [S(\Delta \beta)]^{N_T} \equiv \Big[ \prod_{\gamma=1}^\Gamma e^{- H_{\gamma} \Delta \beta} \Big]^{N_T},
\end{align}
with $\Delta \beta \equiv \frac{\beta}{2N_T}$,
where $H=\sum_{\gamma=1}^\Gamma H_\gamma $ with $H_\gamma$ being each non-commuting term in the $Z_2^{1+1}$ Hamiltonian in \Eq{eq:full_Z2} plus the chemical-potential contribution, after Jordan-Wigner transforming fermionic into spin degrees of freedom. Note that implementing $Q_G$ does not introduce any Trotter error as $[H_n,Q_G]=0$. The Trotter-step ($N_T$) dependence of the chiral condensate $\langle \bar{\Psi} \Psi \rangle$ as a function of chemical potential $\mu$ at a relatively long imaginary time corresponding to temperature $T/m=0.1$ is plotted in \Fig{fig:error_plots}(b). As is seen, with a small number of steps $N_T=10$ ($\Delta \beta = 1$), the chiral phase transition is reproduced qualitatively. To achieve percent-level agreement, $N_T\gtrsim 100$ is required corresponding to step sizes $\Delta \beta \lesssim 0.1$. This suggest that despite large Trotter errors, qualitative features of the phase diagram, such as phase transitions, can be still accessible.

An analytic bound for the multiplicative Trotter error, $M(\Delta \beta) = S(\Delta \beta)/e^{-\Delta \beta \, H}-I$, has been obtained in Ref.~\cite{childs2021theory},
\begin{equation}
||M(\Delta \beta)||=\mathcal{O}(\tilde{\alpha}_{\rm comm}(\Delta \beta)^2 \, e^{4\Delta \beta\sum_{\gamma=1}^{\Gamma} || H_\gamma||}).
\label{eq:trotter_errbound}
\end{equation}
Here, $\tilde{\alpha}_{\rm comm}\equiv \sum_{\gamma_1,\gamma_2=1}^{\Gamma} || [H_{\gamma_2},H_{\gamma_1}]||$ and $||\cdot || $ is the spectral norm. The inset of \Fig{fig:error_plots}(b) shows the multiplicative Trotter error as a function of $\Delta \beta$, along with the numerically determined $||M(\Delta \beta)||$ and the mean trace distance $\bar{D}_{Tr}$ between $e^{-\Delta \beta H}|\Psi_R\rangle$ and $S(\Delta \beta)|\Psi_R\rangle$ averaged over $r=10$ random-state realizations. As can be seen, the numerical results are markedly different from the predicted exponential analytic bound, indicating a power-law dependence on $\Delta \beta$ instead. 
In fact, the multiplicative Trotter error is seen to  scale as $\approx (\Delta \beta)^2$, similar to the scaling of the bound derived in Ref.~\cite{childs2021theory} for Trotterized \textit{real-time} evolution. Indeed, for $e^{-\Delta \beta H}$ with $\Delta \beta >0$ and a physical Hamiltonian $H$ with lower-bounded spectrum, the exponential contribution in \Eq{eq:trotter_errbound} must be replaced by a bounded function to provide a tighter bound.

The influence of device errors on the imaginary-time evolution, through imperfect gates and decoherence, is hardware and algorithm specific and thus difficult to estimate. Nonetheless, one can consider a device-independent approach, in which the errors can be parameterized as imaginary-time evolution under an effective Hamiltonian~\footnote{This may be called `unitary errors' if one were to perform real-time evolution.} $H'=\tilde{H}+H_{\rm err}^\alpha$, with $H_{\rm err}^\alpha$ representing 1- or 2-local  errors with randomized weights bounded by $\alpha$.
Because both fermion (via a Jordan-Wigner transformation) and gauge variables are spin operators, they will be denoted from now on with the same symbol $ \sigma_n^{x/z}, \tilde{\sigma}_n^{x/z}  \rightarrow \sigma_l^{x/z}$, where $l$ denotes sites in a chain of length $2N-1$, and even (odd) sites labeling fermions (gauge fields). 
Motivated by the common bit flip, phase flip, and crosstalk errors in current hardware, we choose $H_{\rm err} = \sum_{l=0}^{2N-1} K_l^{\alpha} \sigma_l$, where $\sigma_l \in\{ \sigma^x_l,\sigma^z_l, \sigma^z_l\sigma^z_{l+1}  \}$, and $K_l^\alpha$ is sampled uniformly from $[-\alpha,\alpha]$. Denoting $|\bar{\Psi} {\Psi} \rangle^{PTPQ}(\alpha)$ the chiral condensate calculated with the PTPQ states under such a noise model, a relative accuracy can be defined as
\begin{align}
A_{\rm rel} \equiv 1-\frac{| \langle  \bar{\Psi} {\Psi} \rangle^{PTPQ}(\alpha) - \langle  \bar{\Psi} {\Psi} \rangle|}{| \langle  \bar{\Psi} {\Psi} \rangle|}.
\end{align}

The chiral condensate under this noise model is plotted in \Fig{fig:error_plots}(c) as a function of chemical potential, at $T/m=0.1$ for $\alpha/m=0.1,0.25,1$, with exact values displayed for reference. For $\sigma^x_l$ errors, PTPQ states are still able to capture phase transitions for $\alpha/m=0.1$, but this ability decays as $\alpha/m \rightarrow 1$, as the relative accuracy near the transition quickly drops to $\sim 45\%$. Furthermore, the inset of \Fig{fig:error_plots}(c) shows $A_{\rm rel}$ for $\mu/m=1.5$ as a function of $\alpha/m$.

A strong difference is observed between $\sigma^z_l$ and $\sigma^x_l$ errors versus $\sigma^z_l\sigma^z_{l+1}$ errors.
$\tilde{\sigma}^x_n$ and $\sigma^z_n$ are both gauge invariant, so the introduced X-type and Z-type noise only violate Gauss's law on half of the qubits. Conversely, ZZ-type noise always violates Gauss's law since each term acts on one link and one site. Since the penalty term $Q_G$ effectively protects against Gauss's-law-violating errors, PTPQ states are seen to be more robust to the latter. In Supplemental Material~\cite{SM}, we discuss the relation between the link-operator convention and the unitary-noise mitigation.

\vspace{0.1 cm}

\noindent
\emph{Scaling and Resource Requirements.} While an exact preparation of Haar-random state has a time complexity that scales exponentially in the system size, pseudo-Haar-random states that are sufficient for most applications can be prepared in polynomial time, as has been established in Refs.~\cite{emerson2003pseudo, oliveira2006efficient, boixo2018characterizing, richter2021simulating}. For example, the (pseudo-) random-state preparation circuit for PTPQ states in this work requires only $\mathcal{O}(Nd)$ two-qubit entangling gates, where $N$ is the number of qubits and $d$ is number of layers. In practice, $d \simeq 20$ is seen to be sufficient in the examples of this work. In general, the required $d$ for given precision depends on the system size but the scaling is observed to be sublinear~\cite{nakata2016efficient,boixo2018characterizing,richter2021simulating}.

To cost the imaginary-time evolution,  we consider the QITE algorithm~\cite{motta2020determining} because of its prominence among near-term approaches. The QITE algorithm consists of Trotterizing imaginary-time evolution, e.g., \Eq{eq:Trotterform}, and it proceeds to find, using classical processing or variationally, a unitary operator $e^{-i \Delta \beta A_\gamma}$ that approximates the non-unitary operator $e^{-H_\gamma \, \Delta\beta}$, such that for given intermediate state $|\Phi \rangle$ in the Trotter sequence,
$e^{-i \Delta \beta A_\gamma} |\Phi \rangle \approx e^{-H_\gamma \, \Delta\beta}  |\Phi \rangle / \mathcal{N}_{\beta}$, where $\mathcal{N}_{\beta} \equiv 
\langle \Phi | e^{-2 H_\gamma \, \Delta\beta}|\Phi \rangle$. Note that for PTPQ-state preparation, $Q_G$ does not need to be Trotterized and can be performed in a single QITE step for time duration $\beta$.

Following this protocol, one incurs two types of errors: the Trotter error $\varepsilon_T \ge || M(\Delta \beta)||$, discussed in the previous section, and the intrinsic QITE error $\varepsilon_Q$ from approximating a non-unitary evolution with a unitary evolution after $N_T$ applications. Here, $A_\gamma$ acts on $N_q$ qubits. Increasing $N_q$ reduces $\varepsilon_Q$ but increases the $A_\gamma$ circuit cost, i.e.,~the total algorithm runtime of the QITE algorithm  is $\Gamma N_T e^{\mathcal{O}(N_q)}$~\cite{motta2020determining,SM}. To estimate this dependence, assume $N_T \geq \beta \eta$ where $ || H_\gamma|| \leq \eta$ is a bound on the operator norm of $H_\gamma$. Then  $ N_q = \mathcal{O} \left(2k\mathcal{C}\ln \left[ {2\sqrt{2}\Gamma N_T}{\varepsilon_Q^{-1}} \right]    \right)$, with $k$ the locality of the Hamiltonian and $\mathcal{C}$ the typical correlation length of the state which is acted upon. The Trotter-step dependence can be exchanged with Trotter error $\varepsilon_T$ using \Eq{eq:trotter_errbound}, but it should be noted that a more favorable behavior is observed empirically, compared to the predicted exponential step-size dependence, see \Fig{fig:error_plots}(b). We add that alternatively, a hybrid interaction-picture QITE-Qubitization simulation protocol can be employed~\cite{Rajput:2021khs} to improve the scaling when the Hamiltonian terms have large spectral norms, such as $Z_2^{1+1}$ in presence of Gauss's law enforcing penalty term.

Note that in essence, the QITE circuit complexity is exponential in the correlation length $\mathcal{C}$. While the PTPQ algorithm starts from an infinite-temperature state, where $\mathcal{C}=0$, this will become prohibitive at a phase transition (of an infinite system), thus highlighting the importance of developing efficient far-term non-unitary evolution schemes. Further details on on the algorithmic scaling is provided in Supplemental Material~\cite{SM}.

\noindent
\textit{Conclusions.}
Physical thermal pure quantum states, introduced in this work as an extension of (canonical) thermal pure quantum states, present  a valuable method for quantum computing thermal phase diagram and thermal non-equal-time correlation functions in strongly interacting gauge theories. Via the example of a $1+1$ dimensional $Z_2$ lattice gauge theory coupled to staggered fermions on a small lattice, PTPQ states are shown numerically to reproduce the phase diagram very accurately over all values of temperature and chemical potential, including a chiral phase transition, with favorable resource requirements. Phase transitions without a local order parameter, such as the (de-)confinement transition present in this model, are strongly finite-size dependent, and will be accessible to future quantum computers. Similarly, thermal no-equal-time correlation functions, important for obtaining transport properties of the Quark-Gluon Plasma in ultra-relativistic heavy ion collisions~\cite{venugopalan2008glasma} can be accessed using the algorithms of this work on the upcoming quantum-computing devices.

\textit{Acknowledgements.} Z.D. and N.M. acknowledge funding by the U.S. Department of Energy’s Office of Science, Office of Nuclear Physics under Award no. DE-SC0021143. Z.D and C.P. acknowledge funding by the U.S. Department of Energy’s Office of Science, Office of Advanced Scientific Computing Research, Accelerated Research in Quantum Computing program award DE-SC0020312. C.P. was further supported by the National Science Foundation under QLCI grant OMA-2120757.

\setcounter{equation}{0}
\setcounter{figure}{0}
\setcounter{table}{0}
\setcounter{page}{1}
\makeatletter
\renewcommand{\theequation}{S\arabic{equation}}
\renewcommand{\thefigure}{S\arabic{figure}}

\section*{Supplementary Material}

\subsection{Thermal Pure Quantum States: 
\\
A Brief Overview
\label{sec:TPQ}}
\noindent
Consider a mechanical observable that can be formally defined by the following bound on its operator norm: $||O|| \leq K N^\ell$, where $K$ is a constant independent of $O$ and $N$, and $\ell \ll N$ denotes the degree of the polynomial of local operators constituting $O$. $N$ denotes the size of the (discrete) system and here for generality, we do not differentiate between the number of lattice sites and the number of qubits encoding system's degrees of freedom. A Thermal Pure Quantum (TPQ) state has the property that the expectation values of all mechanical observables calculated using a TPQ state converge uniformly to those obtained from the ensemble formulation of statistical mechanic in the thermodynamic limit. One can then proceed to find properly-prepared states such that they satisfy this condition.

An ansatz for TPQ states is proposed in Ref.~\cite{sugiura2013canonical},
\begin{equation}
|\beta,N \rangle \equiv e^{-\frac{\beta }{2}H}|\psi_R\rangle\,,
\label{eq:TPQ-SM}
\end{equation}
where $| \psi_R \rangle = \sum_i c_i \ket{i}$ with $\{c_i\}$ being complex numbers drawn randomly from a unit sphere $\sum_i |c_i|^2=1$, and $\{\ket{i}\}$ is any arbitrary orthogonal basis states spanning the full Hilbert space, e.g., a set  of product states in the computational basis.  To show that this ansatz indeed has the property of a TPQ state, one notes that:
\begin{itemize}
\item[i)]{The thermal expectation value in the ensemble formulation is $\braket{O}_{\beta,N}={\text{Tr}(e^{-\beta H}O)}/{\text{Tr}(e^{-\beta H})}$, where trace is defined over any complete orthogonal set of basis states spanning the Hilbert space of a system of size $N$. One then has the identity
\begin{eqnarray}
\llangle  \, \braket{O}_{\beta,N}^{\rm TPQ} \, \rrangle_r
&\equiv& \frac{\llangle  \, \langle \psi_R | e^{-\frac{\beta}{2}H} \,O\, e^{-\frac{\beta}{2}H} | \psi_R \rangle  \, \rrangle_r }{\llangle  \, \langle \psi_R | e^{-\beta H} | \psi_R \rangle  \, \rrangle_r }
\nonumber\\
&&=\frac{\text{Tr}(e^{-\beta H}O)}{\text{Tr}(e^{-\beta H})}
= \braket{O}_{\beta,N},
\end{eqnarray}
in the limit of a sufficiently large number, $r$, of random states. Here, $\llangle \cdot \rrangle_r$ denotes statistical average over $r$ TPQ states, and properties of random states are used to relate the statistical average of expectation values to trace operation~\cite{iitaka2004random}.
}
\item[ii)]{The quantity $D_N^2$, defined by
\begin{equation}
D_N^2 \equiv \llangle  \, \big(\braket{O}^{\rm TPQ}_{\beta,N}-\braket{O}_{\beta,N}\big)^2 \, \rrangle_r\, ,
\label{eq:def-DN2}
\end{equation}
can be shown to be bounded as $D_N^2 \leq N^{2\ell}/e^{\beta\,\Theta(N)}$, see Ref.~\cite{sugiura2013canonical} for a derivation. Here, $\Theta(N)$ indicates that the form is bounded from above and below by $k_1 N$ and $k_2 N$ for two constants $k_1$ and $k_2$ at large $N$.
}
\item[iii)]{Finally, generalized Markov’s inequality puts a bound on the following probability
\begin{eqnarray}
P\big(|\braket{O}_{\beta,N}^{\rm TPQ}-\braket{O}_{\beta,N}| \geq \epsilon\big) \leq \frac{D_N^2}{\epsilon^2} \leq \frac{N^{2\ell}}{\epsilon^2e^{\beta\,\Theta(N)}},
\end{eqnarray}
for any arbitrary $\epsilon>0$. As is seen, this bound exponentially vanishes in the limit of $N \to \infty$, indicating that only $r=1$ random-state realization is sufficient to reproduce, with a high probability, the true ensemble expectation values for large $N$.}
\end{itemize}
This completes the proof that the state defined in Eq.~(\ref{eq:TPQ-SM}) is a TPQ state. Note that for a fixed system size, the higher the temperature, the larger the bound on $D_N^2$, hence requiring more random-state realizations to achieve a given accuracy on thermal expectation values. This feature is evident from the different number of random states used to reproduce, with comparable accuracy, the non-equal-time correlation functions in Fig.~3 of the main text for two different temperatures.

\subsection{Comparison between Standard \\
and Physical TPQ States}
\begin{figure}[t!]
\includegraphics[width=0.8\columnwidth]{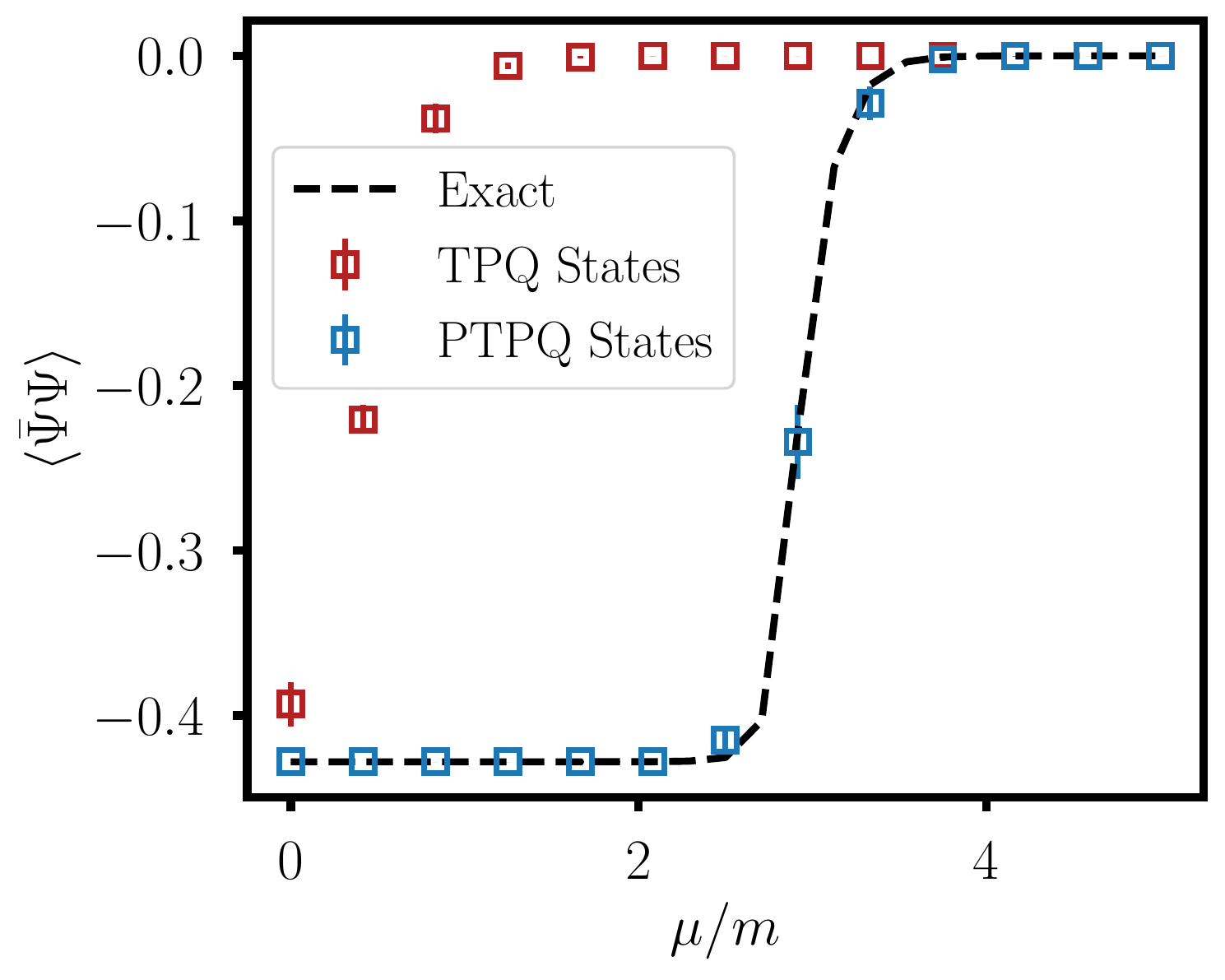}
\caption{Chiral condensate $\langle \bar{\Psi} \Psi \rangle$ at $T/m=0.1$ as a function of (scaled) chemical potential $\mu/m$, calculated with standard TPQ states and PTPQ states. Exact results are also plotted for reference. All results are obtained for $N=6$ fermionic sites, $\lambda/m=4$, $r=10$ random-state realizations, and a random-circuit depth $d=20$.}
\label{fig:standard_comparison}
\end{figure}

\noindent
The need for physical TPQ (PTPQ) states in the context of calculating thermal expectation values of LGTs can be demonstrated by evaluating the chiral condensate of $Z_2^{1+1}$ both with the standard and the physical TPQ states. The result is shown in \Fig{fig:standard_comparison} for $T/m=0.1$, along with the exact values for reference. As in the main text, all results are presented for $m=1/2$, $\epsilon/m=1$ and $a=1$. It is clear that standard TPQ states are not suitable for computing finite-temperature phase diagrams of LGTs, since the errors incurred by drawing the requisite approximate Haar-random state from the entire Hilbert space can lead to incorrect results for the order parameter. This effect is seen to be especially pertinent near the phase transitions. The correct behavior is reproduced by utilizing PTPQ states instead.

\subsection{Chiral Condensate with PTPQ States vs.
\\
Exact Values: Convergence with Lattice Size}
\begin{figure}[t!]
     \centering
     \includegraphics[width=0.9\linewidth]{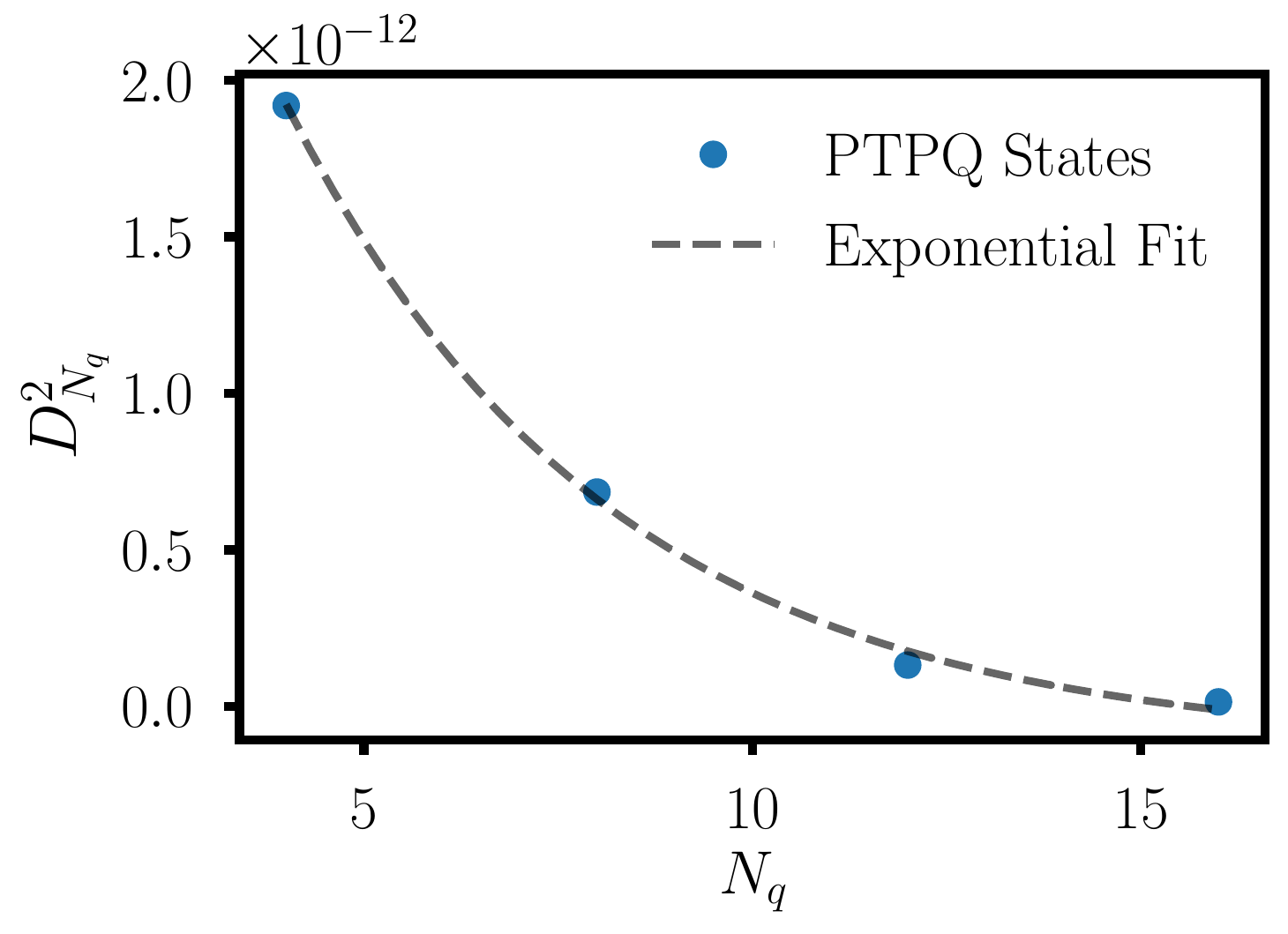}
     \caption{Chiral-condensate error convergence with respect to the number of qubits $N_q$. Results correspond to $T/m=0.1$, $\mu/m=0.2$, and $\lambda/m=100$, averaged over $r=50$ PTPQ states with a random-circuit depth $d=50$.
     }
     \label{fig:error_convergence}
\end{figure}

\noindent
As described in Sec.~\ref{sec:TPQ}, the quantity $D_{N_q}^2$ defined in Eq.~(\ref{eq:def-DN2}) quantifies the convergence of thermal expectation values evaluated with (P)TPQ states to the values obtained within the ensemble formulation of statistical mechanic, and is expected to be bounded by a function that decreases exponentially quickly in the number of qubits $N_q$. This quantity for chiral condensate is plotted in \Fig{fig:error_convergence} as a function of $N_q$. Data from simulated systems up to $N=8$, at $T/m=0.1$, $\mu/m=0.2$, 
and $\lambda/m=100$, averaged over $r=50$ PTPQ states with a random-circuit depth $d=50$, show an exponential decay in $D_{N_q}^2$ consistent with the expectation.

\subsection{Confined Phases of $\mathbb{Z}_2^{1+1}$ and \\ PTPQ States}
\begin{figure}[t!]
\includegraphics[width=0.9\linewidth]{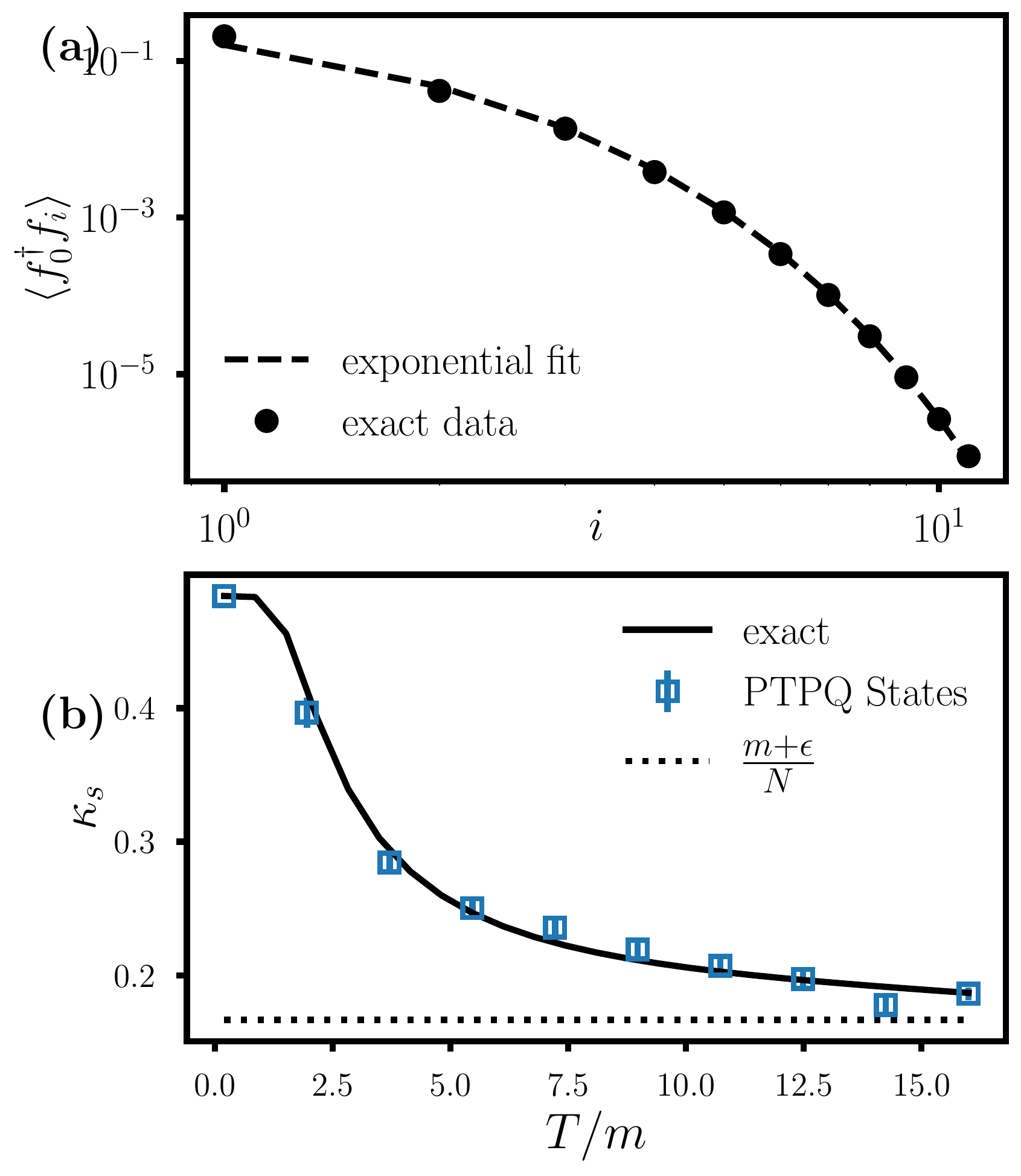}
\caption{(a) Exponential decay of the ground-state two-point fermionic correlator $\langle f_0^\dagger f_i \rangle$ as a function of $i$ for a system of $N=12$ fermionic sites (the Gauss's law is enforced \emph{a priori} hence allowing to numerically simulate larger lattices), indicating a confined phase at $T=0$ and $\mu/m=0.2$. (b) Finite-temperature string tension for a lattice of $N=6$ fermionic sites and at $\mu/m=0.2$, calculated using physical TPQ states with $\lambda/m=30$, $d=20$, and $r=50$. Exact results are plotted for reference.}
\label{fig:gs_and_stringtension}
\end{figure}

\noindent
Determining the finite-temperature confinement-deconfinment phase transition and the associated phase diagram is more challenging due to thermal and finite-size effects, however PTPQ states could still be used to explore these phases with sufficient computational resources. At $T=0$, confinement can be detected by the decay of the non-local two-point correlation function of two gauge-invariant fermionic operators
\begin{equation}
   \langle f^\dagger_0 f_i \rangle~~{\rm with}~~f_i=c_i\bigg[\prod_{j \geq i}\Tilde{\sigma}^z_j\bigg],
\end{equation}
as a function of distance $i$ among them. Explicitly, an exponential decay of this two-point correlator indicates confinement~\cite{borla2020confined}. As depicted in \Fig{fig:gs_and_stringtension}(a), the value of $\langle f^\dagger_0 f_i \rangle$ as a function of $i$ decays exponentially at $\epsilon \neq 0$, where there is expected to be confinement \cite{borla2020confined}. However, a finite temperature can introduce an exponential decay in this correlator even in the absence of confinement, and more robust diagnostics of confinement are required. Additionally, the non-local correlator introduced above constitute a large-order polynomials of local operators, which is prohibitive to the (P)TPQ-state method. 

A better suited quantity for exploring confinement at finite temperatures is string tension, which can be expressed as:
\begin{equation}
    \kappa_s=\frac{E^* - E}{N}.
\end{equation}
Here, $E(E^*)$ is the ground-state energy of the system without (with) a pair of static test charges each placed at one end of the lattice, and $N$ is the number of lattice sites. However, finite-size effects appear to be significant at lattice sizes accessible to our classical exact diagonalization, hindering the identification of confined and deconfined phases. For example, \Fig{fig:gs_and_stringtension}(b) plots string tension for a system of $N=6$ fermionic sites as a function of temperature at $\mu/m=0.2$. In the infinite-temperature limit, where individual local states are purely disordered and $\langle \sigma^x\rangle=\langle\sigma^y\rangle=\langle\sigma^z\rangle= 0$ for all $n$, the system approaches zero energy (dropping constant contributions). Since $\sigma^+_n \sigma^-_n = \frac{\sigma^z_n + I}{2}$, the addition of each test charge yields a mass energy gain of $\frac{m}{2}$, and the test charge at the first lattice site yields an electric field energy gain of $\epsilon$. So, in the infinite temperature limit, the string tension approaches $\frac{m+\epsilon}{N}$. This vanishes in the bulk limit, indicating a deconfined phase at high temperatures. While the numerical values supports convergence onto this value, clearly identifiable regions of confinement and deconfinement outside of this regime would require simulating larger lattices. Nonetheless, TPQ states can still be used to reproduce the expected behavior. Data points corresponding to the string tension evaluated using the PTPQ states are shown in the same plot, indicating good agreement with the exact values. Therefore, PTPQ-state methodology may be used to construct the confinement-deconfinement phase diagram of $\mathbb{Z}_2^{1+1}$ and other LGTs on future quantum computers.

\subsection{Link-Operator Convention and Robustness 
\\
to Unitary Errors}
\noindent
An alternative convention may be used for the link operators such that the electric-field operator is instead represented by $\tilde{\sigma}_n^z$, which switches the roles of $\tilde{\sigma}_n^x$ and $\tilde{\sigma}_n^z$ in the physical Hamiltonian. While this does not affect exact results, the choice of convention does significantly impact how robust PTPQ states are to different types of introduced unitary error. As was seen in the main text, PTPQ states are robust to error terms that violate Gauss's law, as the penalty term in the PTPQ-state preparation effectively suppresses these errors as well. While the common unitary errors in the quantum device are fixed in the computational basis, the operators that violate Gauss's law in each convention are different, hence one expects the most prominent error types to be different for each convention.

\begin{figure}
    \centering
    \includegraphics[width=0.6\linewidth]{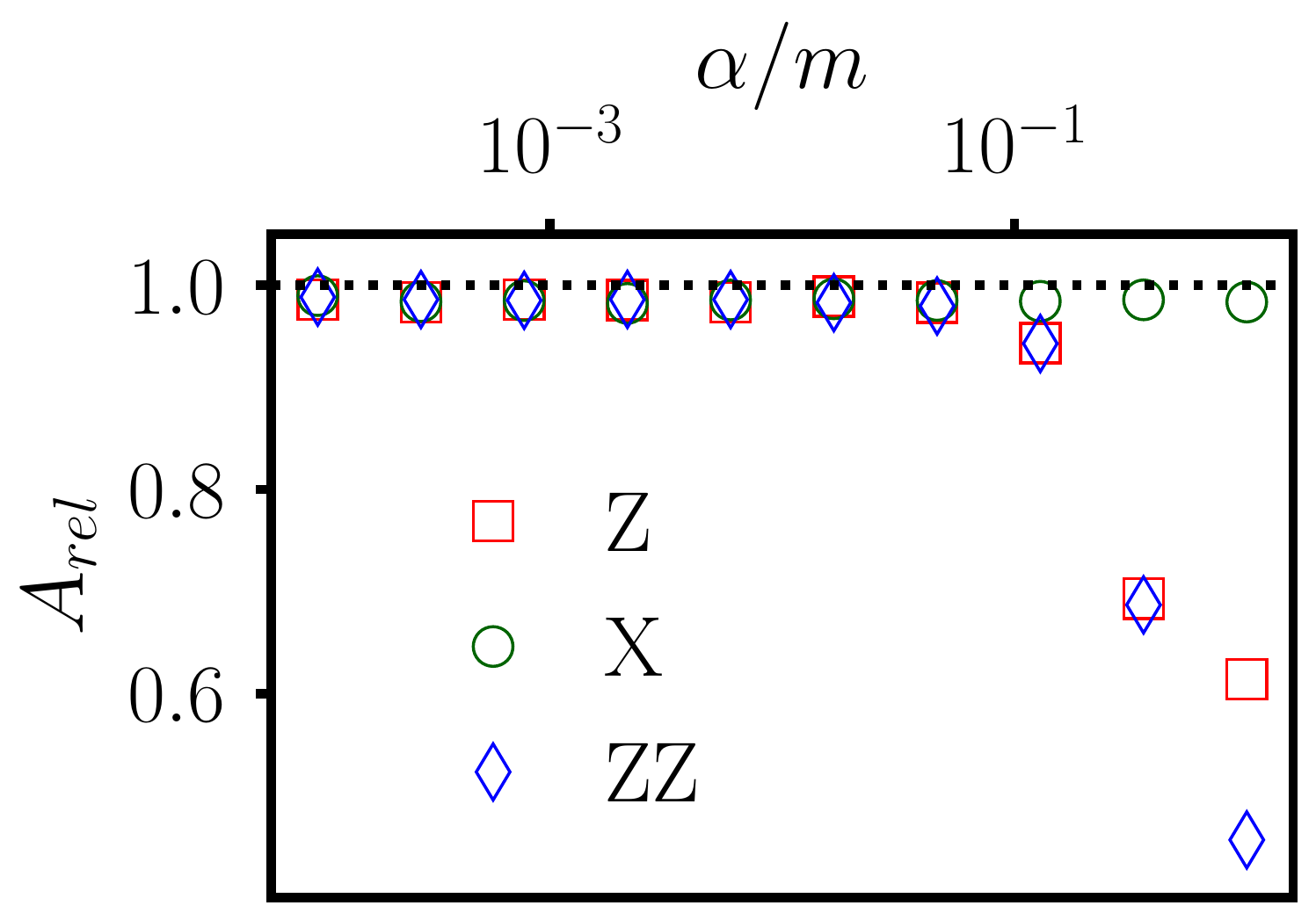}
    \caption{Relative accuracy $A_{\rm rel}$ defined in Eq.~(9) of the main text as a function of relative Hamiltonian error-term strength $\alpha/m$ when various types of unitary error are added to the imaginary-time evolution. All results are obtained for $d=20$, $r=10$, $\lambda/m=8$, $T/m=0.1$, $\mu/m=3$, and averaged over results from 50 randomly generated error terms at each value of $\alpha/m$. Results shown use the alternative convention for link operators, where electric-field operators are represented by $\tilde{\sigma}^z_l$ instead of $\tilde{\sigma}^x_l$.}
    \label{fig:rel_accuracy_old_convention}
\end{figure}

In particular, in the alternate convention to that in the main text, $\sigma^x_n$ and $\tilde{\sigma}^x_n$ both violate Gauss's law, so this type of introduced error is effectively mitigated by $Q_G$. However, neither $\sigma^z_n$ nor $\tilde{\sigma}^z_n$ violate Gauss's law, so PTPQ states are more sensitive to Z and ZZ-type noise. Fig. \ref{fig:rel_accuracy_old_convention} demonstrates this change of sensitivity to the same types of introduced unitary errors as those examined in Fig. 4(c) of the main text. Therefore, the freedom to choose a link-operator convention can be used to increase PTPQ-state-preparation fidelity depending on the most prominent errors seen in a given hardware.

\subsection{Comments on the QITE-algorithm 
\\
Runtime Scaling} 
\noindent
The runtime analysis of the Quantum Imaginary-Time Evolution (QITE) algorithm has been performed in Ref.~\cite{motta2020determining}. The derivation assumes that the norm of each Hamiltonian term is bounded by one---an assumption that can be easily removed upon a slight modification of the results.

Let us start by the reverse triangle inequality,
\begin{equation}
    ||e^{-\Delta \beta H_\gamma}||\geq ||\mathit{I}-\Delta \beta H_\gamma|| \geq 1-\Delta \beta \eta,
\end{equation}
where $|| H_\gamma|| \leq \eta$ for $\eta >0$. Thus, for $N_T=\frac{\beta/2}{\Delta \beta} \geq \beta \eta$~\footnote{Note that the (P)TPQ-state preparation involves evolving with half the inverse temperature, hence the factor of $\beta/2$ here compared with Ref.~\cite{motta2020determining}.}, $||e^{-\Delta \beta H_\gamma}||\geq 1/2$, matching Eq.~(30) of Ref.~\cite{motta2020determining}. Then, following the analysis in Ref.~\cite{motta2020determining} with $N_T\geq \beta \eta$, the total algorithm runtime becomes
\begin{equation}
    \Gamma N_T \, e^{\mathcal{O}\big(k(2\mathcal{C})^d \ln^d(2\sqrt{2}\Gamma N_T\varepsilon_Q^{-1}\big)}.
\label{eq:qiteruntime}
\end{equation}
Here, $\Gamma$ is the number of terms in the Hamiltonian with a given decomposition $H=\sum_{\gamma=1}^\Gamma H_\gamma$, where each Hamiltonian term acts on at most $k$ neighboring qubits on the underlying interaction-to-qubit graph, $\mathcal{C}$ is the upper bound on the correlation length of $|\Phi\rangle$  (in terms of qubits)~\footnote{Correlations between observables separated by distance $L$ are bounded by $e^{- L/\mathcal{C}}$.}, where $\ket{\Phi}$ is a given intermediate state in the Trotter sequence as explained in the main text. Furthermore, $d$ is the spatial dimensionality of the Hamiltonian, and $\varepsilon_Q$ is the error due to approximating non-unitary evolution with a unitary evolution after $N_T$ steps. The total runtime is thus seen to scale quasipolynomially with $\eta$.

Finally, an improvement to the algorithmic scaling of QITE in Ref.~\cite{motta2020determining} is to incorporate the improved product-formula bounds, for example that derived in Ref.~\cite{childs2021theory} combined with the empirical exponential improvement observed in the main text for the physical Hamiltonian of this work. For Hamiltonians of the form $H=\sum_{\gamma=1}^\Gamma H_\gamma$ and 
with first-order product formulas, it follows from Eq.~(43) of Ref.~\cite{childs2021theory} that the Trotter error 
of timestep $\Delta \beta$ is bounded as in Eq.~(8) of the main text. Nonetheless, the exact evaluation of the (multiplicative) error for small system sizes reveals that the error in each Trotter step goes as $\mathcal{O}((\Delta \beta)^2)$, hence the number of Trotter steps to reach an overall accuracy $\varepsilon_T$ after evolving for $\beta/2=N_T \Delta \beta$ in (P)TPQ-state preparation goes as $N_T=\mathcal{O}(\beta^2/\varepsilon_T)$. Merging this with the QITE resource requirements discussed in the main text and in Ref.~\cite{motta2020determining}, provided that $N_T \geq \beta \eta$, one concludes that each unitary solved for within the QITE algorithm must act on at most $N_q$ number of qubits as given in the main text.


\bibliography{references_cleaned}

\end{document}